\documentclass[prb,aps,superscriptaddress,twocolumn,showpacs,preprintnumbers,amsmath,amssymb,reprint]{revtex4-1}

\usepackage{graphicx}
\usepackage{amsmath}
\usepackage{amssymb}
\usepackage{color}
\usepackage{float}
\usepackage{dcolumn}
\usepackage{epsfig}
\usepackage{bm}
\usepackage{subfigure}
\usepackage[urlcolor=blue]{hyperref}
\hypersetup{backref, colorlinks=true, linkcolor=blue, citecolor=blue}
\begin{document}

\title{Transport evidence for the surface state and spin-phonon interaction in FeTe$_{0.5}$Se$_{0.5}$}

\author{Mu-Yun Li }
\thanks{These authors contributed equally to this work.}
\affiliation{Center for High Pressure Science and Technology Advanced Research, Shanghai 201203, China}
\affiliation{Shanghai Institute of Space Power Source, Shanghai 200245, China}

\author{Jia-Wei Hu}
\thanks{These authors contributed equally to this work.}
\affiliation{Key Laboratory of Materials Physics, Institute of Solid State Physics, HFIPS, Chinese Academy of Sciences, Hefei 230031, China}
\affiliation{University of Science and Technology of China, Hefei 230026, China}

\author{Ge Huang}
\affiliation{Center for High Pressure Science and Technology Advanced Research, Shanghai 201203, China}

\author{Wei-Jian Li}
\affiliation{Center for High Pressure Science and Technology Advanced Research, Shanghai 201203, China}

\author{Ya-Kang Peng}
\affiliation{Center for High Pressure Science and Technology Advanced Research, Shanghai 201203, China}

\author{Guangyong Xu}
\affiliation{NIST Center for Neutron Research, National Institute of Standards and Technology,  Gaithersburg, MD 20899, USA}

\author{Genda Gu}
\affiliation{Condensed Matter Physics \& Materials Science Division, Brookhaven National Laboratory, Upton, NY 11973, USA}

\author{Xiao-Jia Chen}
\email{xjchen2@gmail.com}
\affiliation{Center for High Pressure Science and Technology Advanced Research, Shanghai 201203, China}
\affiliation{School of Science, Harbin Institute of Technology, Shenzhen 518055, China}

\date{\today}

\begin{abstract}

The iron chalcogenides have been proved to be intrinsic topological superconductors to implement quantum computation because of their unique electronic structures. The topologically nontrivial surface states of FeTe$_{0.5}$Se$_{0.5}$ have been predicted by several calculations and then confirmed by high-resolution photoemission and scanning tunneling experiments. However, so far, the shreds of the electrical transport evidence for topological surface states are still in absence. By carrying out electrical transport experiments, we observe a topological transition with a nonlinear Hall conductivity and simultaneous linear magnetoresistance near the superconducting transition temperature. Furthermore, we observe a sign reversal of the Hall coefficient accompanied by a concurrently softening of the ${A}_{1g}$ phonon mode at about 40 K, indicating a nematic transition. The synchronized phonon softening with nematicity manifests an enhanced fluctuation state through spin-phonon interaction. Our results solidly corroborate the topological surface states of FeTe$_{0.5}$Se$_{0.5}$ and provide an understanding of the mechanism of the superconductivity in iron chalcogenides.

\end{abstract}

\maketitle

\section{Introduction}
The iron chalcogenides are one of the most favored materials in the field of topological superconductivity\cite{xu} because of their structural simplicity\cite{1}, uncomplicated phase diagram\cite{2}, and easily-substitution-tuned physical properties\cite{3,4}. The parent compound FeSe has a transition temperature (${T}_{c}$) of 8 K under ambient pressure\cite{5}, and of 37 K at a pressure of about 9 GPa\cite{6}. Through proximity effect, a surprisingly higher ${T}_{c}$ of 65 K was reported accompanied by a topological phase in the single-layer FeSe film on SrTiO$_{3}$\cite{7}. It has been proposed that the lattice vibrations\cite{8} and excitations of electronic origin such as spin or electronic polarizability fluctuations\cite{5} are the two important candidates of Cooper pairing for high-$T_c$ superconductivity in these superconductors. Later well-designed studies revealed that both are important to the pairing interaction strengths in FeSe\cite{9}. The strong enhancement of $T_{c}$ of the monolayer FeSe film was suggested to result from the strong electron-phonon coupling (EPC) played by an interface soft mode\cite{9,10,66}. Therefore, identifying the potential soft phonon modes in unconventional superconductors remains an arduous but necessary task and will help to understand the mechanism of superconductivity\cite{70}. With increasing the concentration of the Te atom, an intense spin-orbit coupling effect is introduced in Fe$_{1+y}$T$e_{x}$Se$_{1-x}$ systems\cite{20} and the ${T}_{c}$ rises to 14.5 K at $x$ around 0.5\cite{11}. The spin-orbit interaction will cause a band inversion near the Brillouin center. Thus, an $s$-wave superconducting gap on the surface state of FeTe$_{0.5}$Se$_{0.5}$ below ${T}_{c}$ was observed by the recent angle-resolved photoemission spectroscopy (ARPES) experiments, confirming the existence of the topological superconductivity\cite{11,12}. By tuning the nontrivial superconductivity on the surface, the iron chalcogenides can be a promising candidate for realizing Majorana bound states and paving the way for quantum computation\cite{57,13,14}.  

The inverted electronic structure and spin-helical surface texture for FeTe$_{0.5}$Se$_{0.5}$ resemble those of topological insulators\cite{11,12,58}.
As reported by Qu $et$ $al$, the acquired transport parameters of surface states can help to investigate the high-mobility Dirac fermions and Majorana bound states\cite{19}.
So far, there is little information on FeTe$_{0.5}$Se$_{0.5}$ in this aspect\cite{12}. The lack of transport information on surface state, especially the carrier mobility, has hindered the further understanding of topological states in FeTe$_{0.5}$Se$_{0.5}$. Meanwhile, the mechanisms of unconventional superconductivity are still controversial issues\cite{5,8,9}. Previous Raman scattering experiments on iron chalcogenides were mainly focused on the lattice dynamics\cite{15,16,17,18}, while the features for superconductivity or other electronic states such as nematicity have not been reliably clarified. The systematical studies on the vibrational properties of high-quality single-crystal FeTe$_{0.5}$Se$_{0.5}$ are desirable. This could help the understanding of the mechanism(s) of superconductivity and the possible role of the electron-phonon coupling in iron-based superconductors. 

In this work, we carry out the first-principles calculations, temperature-dependent electrical transport, and Raman scattering measurements to explore the topological superconductivity of single-crystal FeTe$_{0.5}$Se$_{0.5}$. We corroborate a topological surface state with high mobility through transport measurements, which is consistent with our theoretical calculations. Combined with vibrational properties, we generally demonstrate the phase diagram of FeTe$_{0.5}$Se$_{0.5}$. We find a nematic transition followed by the appearance of the topological surface state. We conclude that both the spins and phonons play an essential role in the unconventional superconductivity of iron chalcogenides.

\begin{figure}[t]
	\includegraphics[width=1.08\columnwidth]{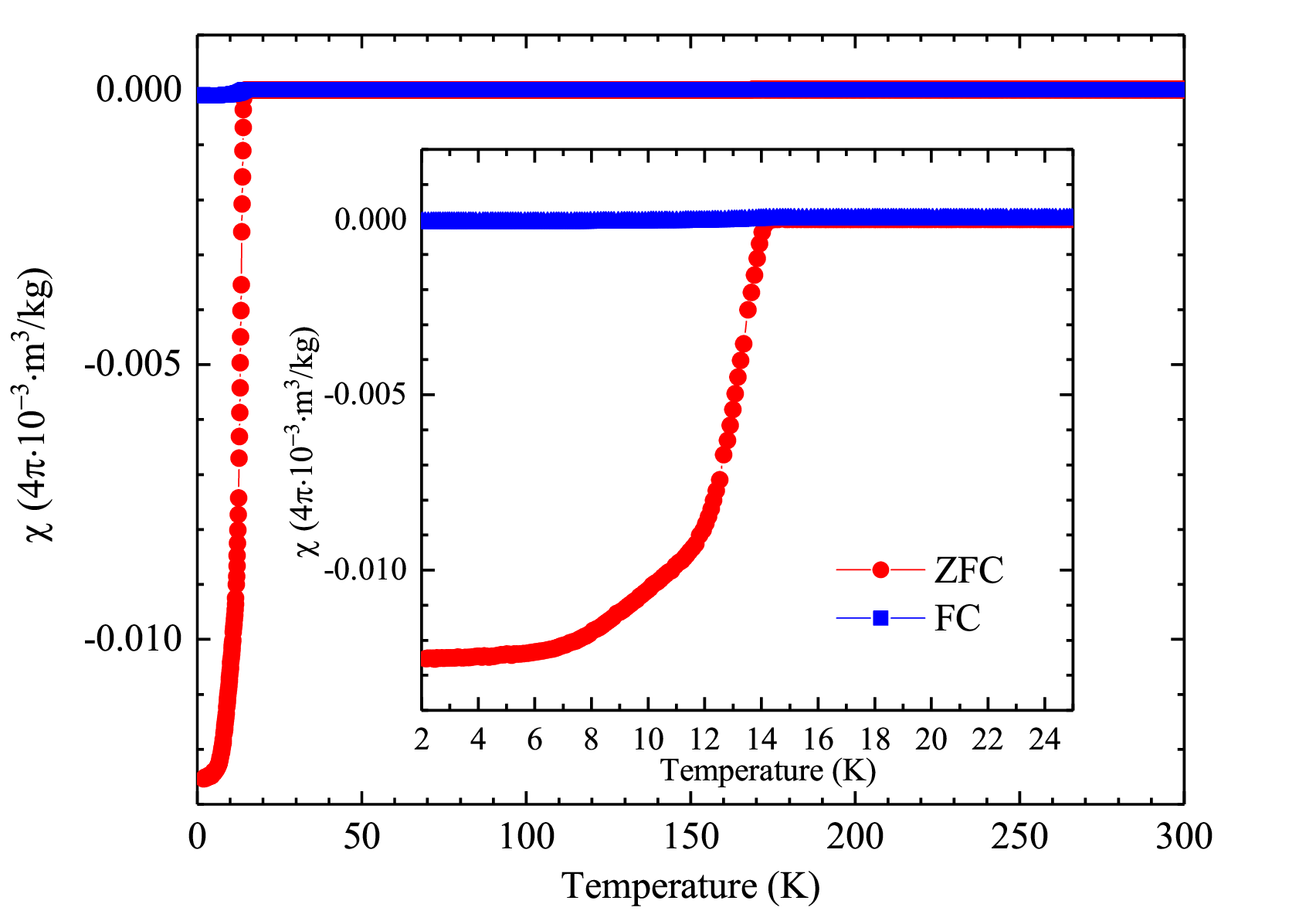}
	\setlength{\abovecaptionskip}{0.cm}
	\caption{Temperature dependence of the dc magnetization of FeTe$_{0.5}$Se$_{0.5}$. The magnetic fields were set to be 20 Oe. The red and blue symbols denote the data points of zero-field cooling (ZFC) and field cooling (FC), respectively. The inset demonstrates the magnification in the temperature range from 2 to 25 K. The ${T}_{c}$ was determined to be 14.5 K for the studied sample.}
\end{figure}

\begin{figure*}[t!]
	\includegraphics[width=2\columnwidth]{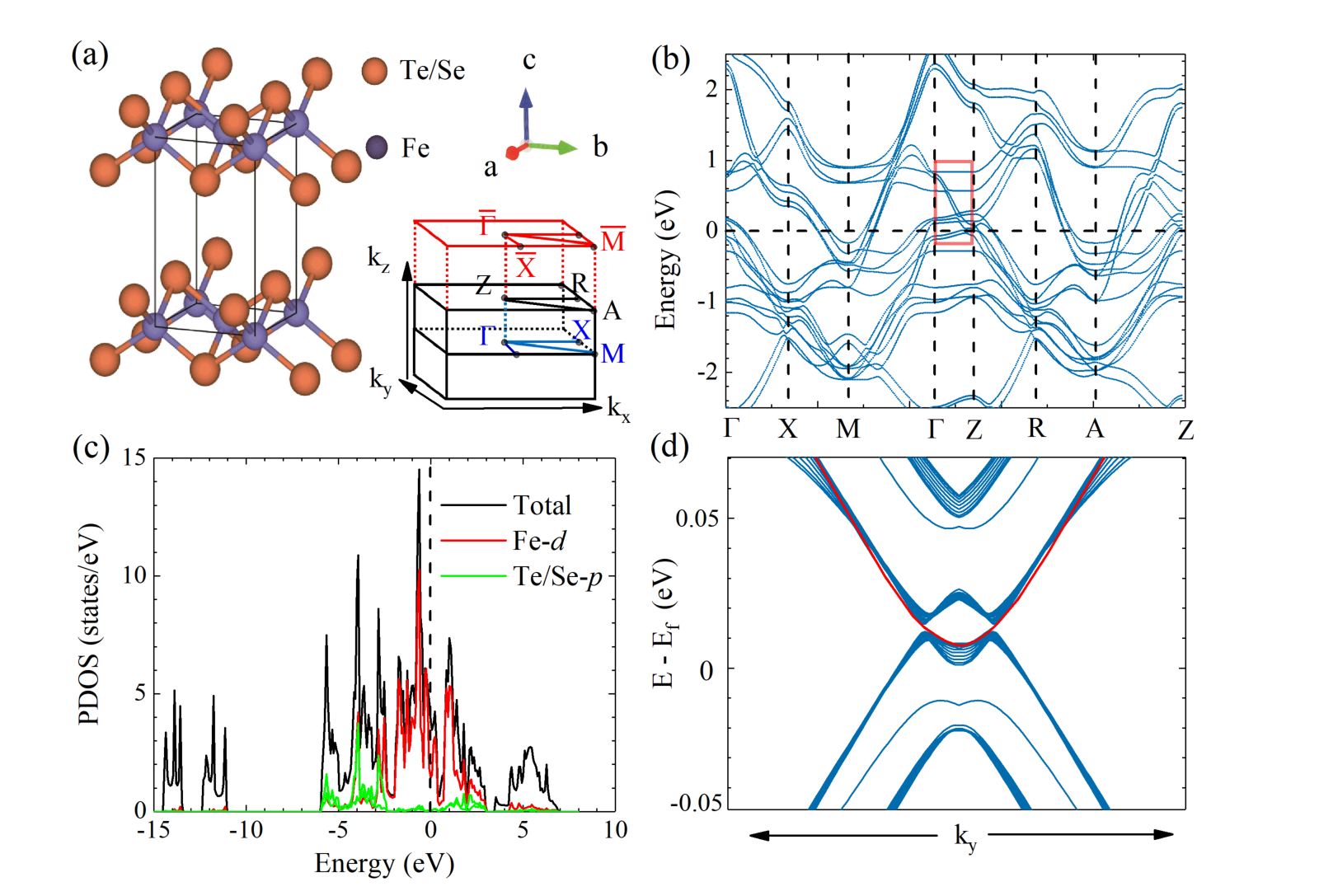}
	\setlength{\abovecaptionskip}{0.cm}
	\caption{(a) The schematic crystalline structure of Fe(Te,Se) and the sketch of the Brillouin zone of FeTe$_{0.5}$Se$_{0.5}$. The in-plane (001) surface Brillouin zone is projected along the ${k}_{z}$ direction and marked by red solid lines. (b) Electronic band structure of FeTe$_{0.5}$Se$_{0.5}$. The red line marks the area where the band inversion emerges. (c) Electronic density of states of FeTe$_{0.5}$Se$_{0.5}$. The black solid line represents the total density of states. The red and green solid line denote the projected density of states of the Fe ($d$-orbit) and Te/Se ($p$-orbit) atoms, respectively. (d) Electronic band structure projected to the (001) surface. The red solid line shows the evident nontrivial Dirac-cone state.}
\end{figure*}

\section{Experimental and Calculation Details}

High-quality single crystal with nominal composition FeTe$_{0.5}$Se$_{0.5}$ was grown by the unidirectional solidification method. The oxygen-annealing method was used to remove the excess Fe atoms in the interstitial sites. The ${T}_{c}$ of 14.5 K was determined by magnetization measurements with Quantum Design's Magnetic Properties Measurement System, as shown in Fig. 1. 
Fresh and clean surface of FeTe$_{0.5}$Se$_{0.5}$ was cleaved in the glove box for electrical transport and Raman scattering measurements with the size of about 150 $\times$ 150 $\mu$m and the thickness of about 25 $\mu$m.

The electrical resistivity (${\rho}$) was measured with Quantum Design's Physical Property Measurement System using the standard four-probe method and Platinum electrodes.
To eliminate the transverse (longitudinal) resistivity component from the misalignment of contacts, we obtained the ${\rho}_{xy}$ and ${\rho}_{xx}$ through ${\rho}_{xy}=[{\rho}_{xy}{(+H)}-{\rho}_{xy}{(-H)}]/2$ and ${\rho}_{xx}=[{\rho}_{xx}{(+H)}+{\rho}_{xx}{(-H)}]/2$, respectively. The components for the Hall conductivity along the $xx$ and $xy$ directions are related to the two expressions $\sigma_{xx}=1/\rho_{xx}$ and $\sigma_{xy}=\rho_{xy}/(\rho_{xx}^{2}+\rho_{xy}^{2})$.

The Raman spectra were collected using a single-stage spectrograph equipped with a thermoelectrically cooled charge-coupled device. The measurements were carried out on a freshly cleaved sample surface of the crystal using a laser with the wavelength of 488 nm in the temperature range from 296 to 5 K. The incident power was set below 1.5 mW with a spot diameter of about 10 $\mu$m to avoid the heating effect. 

All the first-principles calculations were performed in the framework of the density functional theory\cite{48,49}, as implemented in the Vienna $ab$ $inito$ simulation package\cite{50,51,52}. The spin-orbit coupling was included in the calculations. The Perdew-Burke-Ernzerhof of generalized gradient approximation was used for the exchange-correlation potential\cite{53}. The optimized experimental lattice parameters\cite{54} were used to construct the crystal structure. The unit cell with nominal composition FeTe$_{0.5}$Se$_{0.5}$ was extended to a large one in the $c$ direction, with two Fe atoms and one atom for each of the Te and Se elements involved (Fe$_{2}$TeSe). For both self-consistent field and non-self-consistent field calculations, a 9$\times$9$\times$6 $k$-point mesh was used in the Brillouin zone, and the cut-off energy was set to 520 eV. The open-source software package (Wanniertools)\cite{55} was used to investigate the topological nature on the projected (001) surface of FeTe$_{0.5}$Se$_{0.5}$. The $s$ and $p$ orbits for Fe atoms and the $p$ orbits for Te/Se atoms were selected when employing the maximally localized Wannier functions\cite{56}.

\section{RESULTS AND DISCUSSION}

\subsection{Dirac-cone-like surface state from calculations}

The schematic structure of layered FeTe$_{x}$Se$_{1-x}$ with ordered Te/Se sites and the sketched Brillouin zone of FeTe$_{0.5}$Se$_{0.5}$ are presented in Fig. 2(a). Each of the Fe atoms is connected with four neighboring Te/Se atoms, forming a tetrahedron. Tunning the dopant level of Te/Se atoms introduces an enormous spin-orbit coupling effect and significantly influences the band structure near Fermi level (${E}_{f}$)\cite{11}. A topological transition emerges at around $x$ = 0.5\cite{11,12}. The Brillouin zone of FeTe$_{0.5}$Se$_{0.5}$ has eight points with time-reversal symmetry which include $\Gamma$(0,0,0), $M$($\pi$,$\pi$,0), $Z$(0,0,$\pi$), $A$($\pi$,$\pi$,$\pi$), two $X$($\pi$,0,0), and two $R$($\pi$,0,$\pi$). The two-dimensional Brillouin zone of the (001) surface is projected along the ${k}_{z}$ direction. Based on the $k$ points of the Brillouin zone, Fig. 2(b) gives the calculated bulk electronic structure of FeTe$_{0.5}$Se$_{0.5}$. The band dispersion of the bulk state is generally consistent with the preceding theoretical predictions and ARPES experiments\cite{20,11,12}. Previous studies substantiated the Dirac-cone-like surface states of FeTe$_{0.5}$Se$_{0.5}$ in the proximity to bulk superconductivity\cite{11,12}. Within the spin-orbit coupling, the nontrivial topological superconductivity is induced by the band inversion between the Fe-$d$ and Te/Se-$p$ orbits along the $\Gamma$ to $Z$ direction. The place where the band inversion appears is marked by the red frame. For a better interpretation, Fig. 2(c) demonstrates the density of states for different atoms. The Fe-$d$ and Te/Se-$p$ orbits dominate the Fermi surface topology near ${E}_{f}$, although the weight of Te/Se-$p$ components is relatively small.  Based on the reliable bulk calculations, we constructed the tight-binding model to reveal the (001) surface's topological nature. Figure 2(d) displays the calculated Dirac-cone-like surface state along the ${k}_{y}$ direction. An evident linear dispersion of the band structure near the ${E}_{f}$ can be observed. The calculations demonstrate that the surface state is gapless and topologically nontrivial, consistent with previous reports\cite{11,12}. The linear dispersion of the electronic structure shares similarities with the surface-state band structures in topological insulators\cite{19}. From the calculations, we conclude that FeTe$_{0.5}$Se$_{0.5}$ is a topological superconductor. The unique electronic structure indicates that when the neighboring bands are twisted into the Dirac cones, the carriers become high-mobility Dirac fermions. The results of theoretical calculations provide insightful guidance for the following experiments.

\subsection{Electrical transport properties}

\begin{figure*}[htp]
	\includegraphics[width=2\columnwidth]{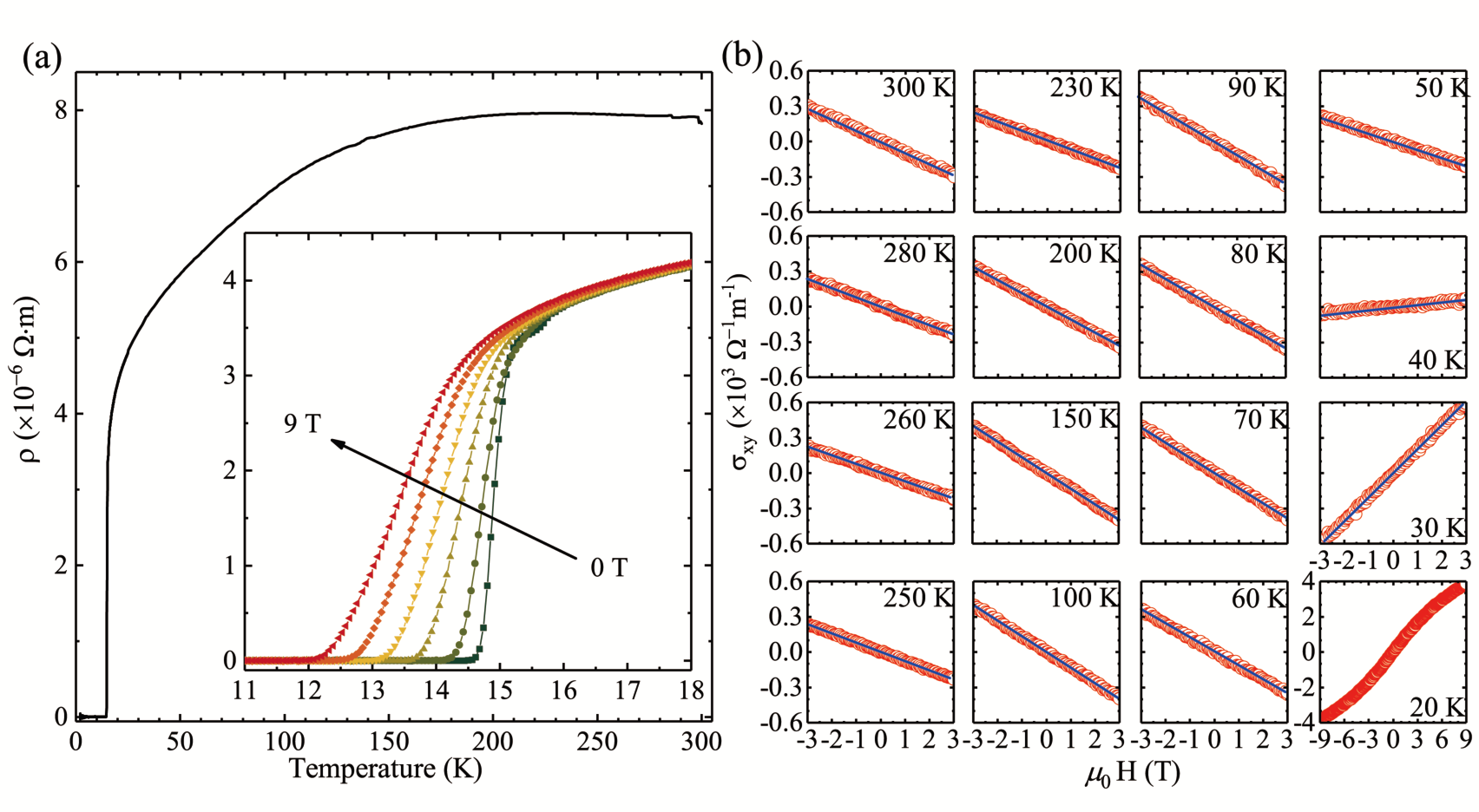}
	\setlength{\abovecaptionskip}{0.cm}
	\caption{(a) Temperature dependence of the electrical resistivity of FeTe$_{0.5}$Se$_{0.5}$ single crystal. The inset shows the temperature dependence of the resistivity at various magnetic fields up to 9 T. The applied field is parallel to the $c$ axis. (b) Representative Hall conductivities of  FeTe$_{0.5}$Se$_{0.5}$ at different temperatures. The red open circles and blue solid lines denote the experimental data points and linear fitting curves, respectively. The magnetic fields were set to change from -9 T to 9 T for a temperature of 20 K. At other temperatures, the magnetic fields were set to change from -3 T to 3 T.}
\end{figure*}

Before carrying out the Hall effect measurements, it is necessary to examine the quality of the sample. Figure 3(a) presents the temperature dependence of the resistivity for FeTe$_{0.5}$Se$_{0.5}$. The inset displays the resistivities below 18 K at each applied magnetic field from 0 T to 9 T. The sample turns into a metallic state below 250 K ($d\rho/dT>0$) and exhibits superconductivity at the temperature of about 15 K [See the magnetization measurements shown in the Fig. 1(a)]. With increasing magnetic field, the resistivity transition broadens and the ${T}_{c}$ shifts to lower temperatures. Some similar broadening behaviors have been reported previously in other 11-type Fe-based systems\cite{21}. The relative high ${T}_{c}$ value and metallic resistivity without the antiferromagnetic-like upturn illustrate that the crystal used is well annealed and that the distribution of excess Fe interstitials can be ruled out\cite{22,23}. The excess Fe usually comes from inadequate O$_{2}$-annealing and significantly affects the electronic properties. It could provide extra moments that localize the electrons\cite{24,25} and even cause a negative magnetoresistance\cite{22}. The excellent growth quality of our sample makes us confident for the following experiments and the related results. 

For studying the topological properties of FeTe$_{0.5}$Se$_{0.5}$, we performed temperature-dependent Hall resistivity measurements. Figure 3(b) shows the Hall conductivity ($\sigma_{xy}$) at applied magnetic fields for some representative temperatures. The obtained Hall conductivity $\sigma_{xy}$ follows a linear relationship with the applied magnetic field up to 3 T for temperatures from 300 K down to 30 K. Thus, the Hall coefficient (${R}_{H}$) and carrier concentration (${n}_{H}$) can be obtained from the single-band model\cite{26}: 
\begin{equation}
R_{H}=\rho_{xy}/\mu_{0}H~~.
\end{equation}
\begin{equation}
n_{H}=1/eR_{H}~~.
\end{equation}
The slope ($d\sigma_{xy}/dH$) keeps negative above 50 K and becomes positive below 40 K, indicating that the dominant carriers change from the electrons to holes. The transport properties of dominant hole-like carriers at low temperature are similar to those reported by Pimentel $et$ $al$\cite{29} but differ from others\cite{27,28}. In any case, as a common phenomenon in iron chalcogenides, the sign reversal of ${R}_{H}$ near 40 K can be explained by the Fermi surface reconstruction that leads to the nematic transition\cite{46}. In particular, numerous ARPES experiments reported the splitting of ${d}_{xz}/{d}_{yz}$ bands near ${E}_{f}$, which triggers the structure evolution from the four-fold to two-fold symmetry\cite{30,31,32,28}.

When the temperature is lowered to 20 K, a distinctive nonlinear behavior is observed. Two features are noteworthy for the Hall conductivity at 20 K: Firstly, the $\sigma_{xy}$ (or $\rho_{xy}$) deviates from linearity in the low field range, while in the normal state above 30 K, the linear behavior maintains up to 3 T. Secondly, the slope of the $\sigma_{xy}$-$H$ curve remains positive up to 9 T, again confirming the dominant hole carriers in the low-temperature transport within the whole field range. For the topological insulating state, the Hall resistivity does not exhibit a linear behavior but has a resonant structure at the small magnetic field. The transport anomalies at 20 K are similar to those in other topological insulators, such as the Bi$_2$Te$_{3}$ and Bi$_{2}$Se$_{3}$\cite{19,64}, in which the surface states play a non-negligible role in the overall transport processes. Considering the recent ARPES results for FeTe$_{0.5}$Se$_{0.5}$ with a topological-insulator Dirac surface state at 15 K\cite{11,12}, we believe that the Hall and MR anomalies appearing in the vicinity of superconductivity are induced by the nontrivial electronic structure. Furthermore, a linear dependence of the magnetoresistance (MR) on the magnetic field appears coincidently, as shown in Fig. 4(b), which can be interpreted as the result of the Dirac cone states\cite{34,12,64}. The results mentioned above offer an excellent opportunity for us to investigate the contributions of the surface state to the transport properties.

\subsection{Transport evidence for the surface state}

\begin{figure}[t!]
	\includegraphics[width=1\columnwidth]{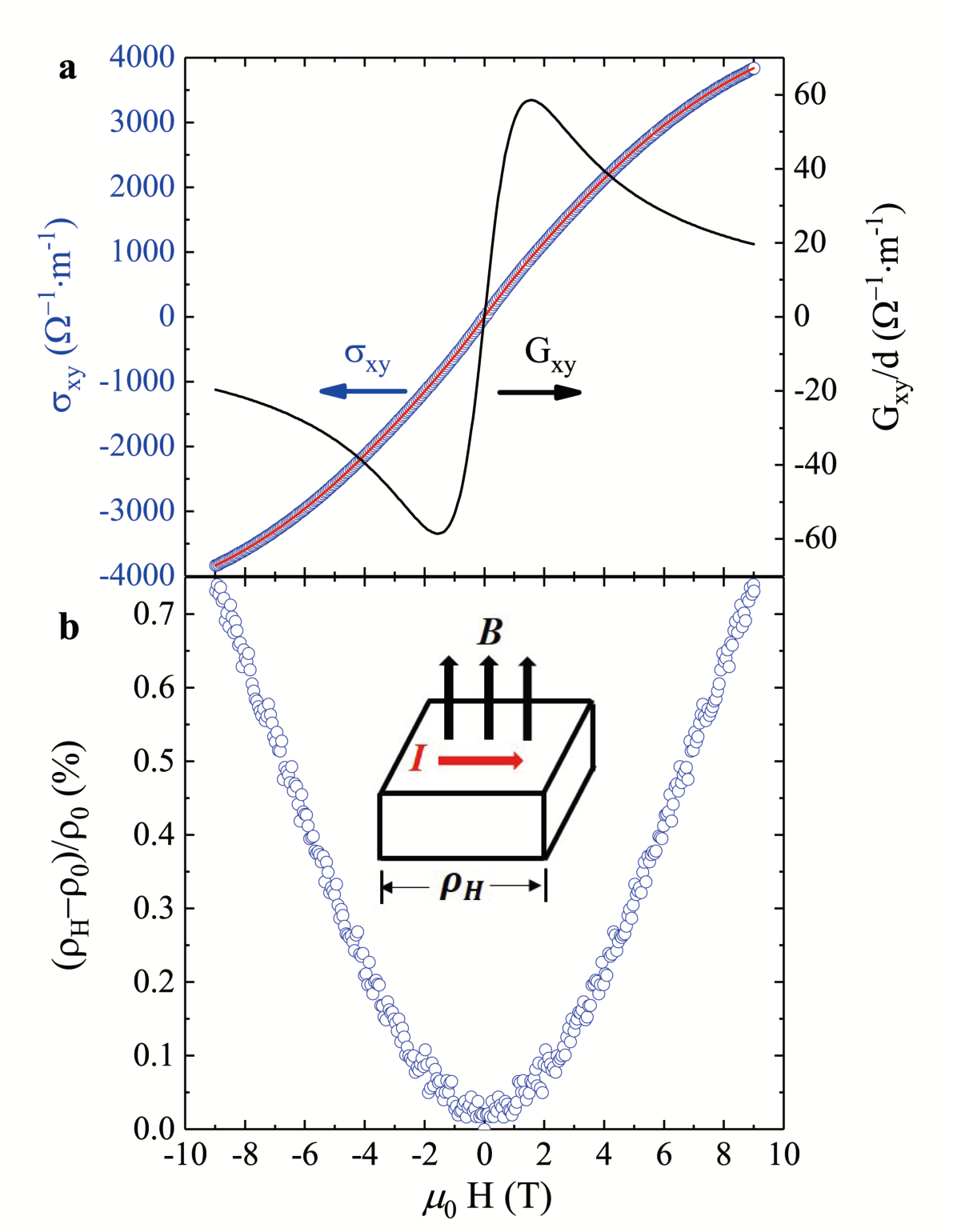}
	\setlength{\abovecaptionskip}{0.cm}
	\caption{(a) Magnetic field dependence of the Hall conductivity of FeTe$_{0.5}$Se$_{0.5}$ at the temperature of 20 K up to 9 T. The blue arrow pointing to the left axis denotes the whole conductivity ($\sigma_{xy}$). The black arrow pointing to the right axis represents the relevant surface term ($G_{xy}$). The blue circles denote the raw data points, with the red solid lines denoting the whole fitting results. (b)  The magnetic-field-dependent magnetoresistance [($\rho_{H}-\rho_{0}$)/$\rho_{0}$] of  FeTe$_{0.5}$Se$_{0.5}$ at the temperature of 20 K for the applied magnetic fields up to 9 T. The inset shows the schematic picture of measurements.}
\end{figure}

\begin{figure}[htb]
	\includegraphics[width=1\columnwidth]{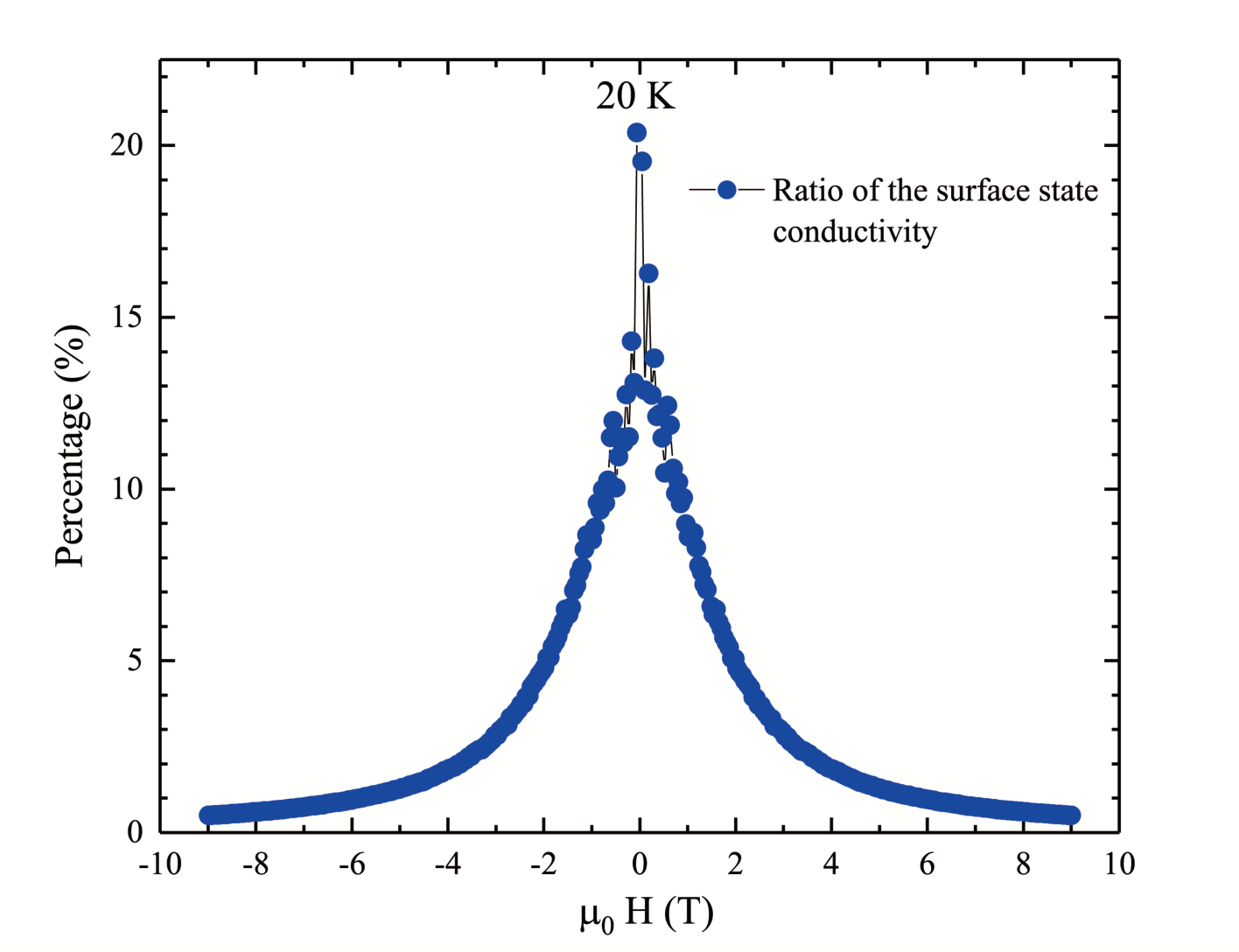}
	\setlength{\abovecaptionskip}{0.cm}
	\caption{The contribution of the surface state in FeTe$_{0.5}$Se$_{0.5}$ to the total conductivity at the magnetic field from -9 T to 9 T. The blue symbols denote the ratio of the surface state conductivity.}
\end{figure}

Figure 4 displays the Hall conductivity and MR at a temperature of 20 K as a function of the magnetic fields up to 9 T. The MR exhibits conventional parabolic lineshape at low magnetic fields. At higher fields, the MR increases linearly due to the quantum limit and corroborates the Dirac-cone-like band dispersion\cite{34,11,12}. The behavior of MR is a bit different in topological insulators, where the linear MR was also identified in the low magnetic field range\cite{19,35}. This is because the topological insulators have a much larger Landau level splitting than the conventional bulk bands\cite{34}. Therefore, the quantum limit of topological insulators can be achieved in the lower magnetic field regions\cite{36}. The MR of FeTe$_{0.5}$Se$_{0.5}$ at 9 T and 20 K (0.7\%) is much larger than those of as-grown (0.03\%) and half-annealed (0.14\%) samples of Fe$_{1+y}$Te$_{0.6}$Se$_{0.4}$\cite{34}.  The relatively large and positive MR again demonstrates the good growth quality of our sample\cite{22}. The observed Hall conductivity $\sigma_{xy}$ can be described as the sum of the surface Hall conductivity $\sigma_{xy}^s$ and the bulk $\sigma_{xy}^b$\cite{19}:
\begin{equation}
\sigma_{xy}=\sigma_{xy}^{s}+\sigma_{xy}^{b}~~.
\end{equation}
The surface conductivity $\sigma_{xy}^s$ can be expressed as:
\begin{equation}
\sigma_{xy}^{s}=G_{xy}/d=\frac{2\pi e^{3}}{dh^{2}}\frac{Bl^{2}}{1+(\mu_{s}B)^2}~~.
\end{equation}
Where $B$ is the magnetic flux density, $d$ is the sample thickness, $l$ is the mean free path, and $\mu_s$ is the surface carrier mobility. The bulk conductivity $\sigma_{xy}^b$ can be expressed by the semiclassical form:
\begin{equation}
\sigma_{xy}^{b}=p_{eff}e\mu_{b}\frac{\mu_{b}B}{1+(\mu_{b}B)^2}~~.
\end{equation}
Where $p_{eff}$ is the effective carrier concentration and $\mu_{b}$ is the bulk carrier mobility. The fitting matches well with the experimental data and gives $P_{eff}$ = 1.65$\times$ 10${{}^{23}}$ m$^{-3}$ and $l$ = 179 nm. The acquired value of the mean path for FeTe$_{0.5}$Se$_{0.5}$ is larger than that of FeSe (30 nm for the outer hole pocket $\beta$ and 80 nm for the inner hole pocket $\delta$ in the Brillouin center)\cite{37}, but close to that of Bi$_{2}$Se$_{3}$ (235 nm)\cite{19}. Intriguingly, from Eqs. (3)-(5), the obtained mobility of the surface state (6440 cm$^{2}$V$^{-1}$s$^{-1}$) is almost ten times larger than that of the bulk state (610 cm$^{2}$V$^{-1}$s$^{-1}$). To be more specific, Fig. 5 shows the ratio of the surface state conductivity to the total conductivity ($\sigma_{xy}^{s}/\sigma_{xy}$). In the low magnetic field range, the surface state accounts for about 5\% to a maximum of 20\%. Despite the relatively small weight, the surface state can dominate the low-magnetic-field electrical transport with much higher mobility induced by the Dirac-cone-like band dispersion. The resonance Hall conductivity and nearly linear magnetoresistance have also been observed to serve as the evidence of nontrivial topological state in highly efficient thermoelectric materials\cite{lcc,pang}.  

\begin{figure}[tbp]
	\includegraphics[width=\columnwidth]{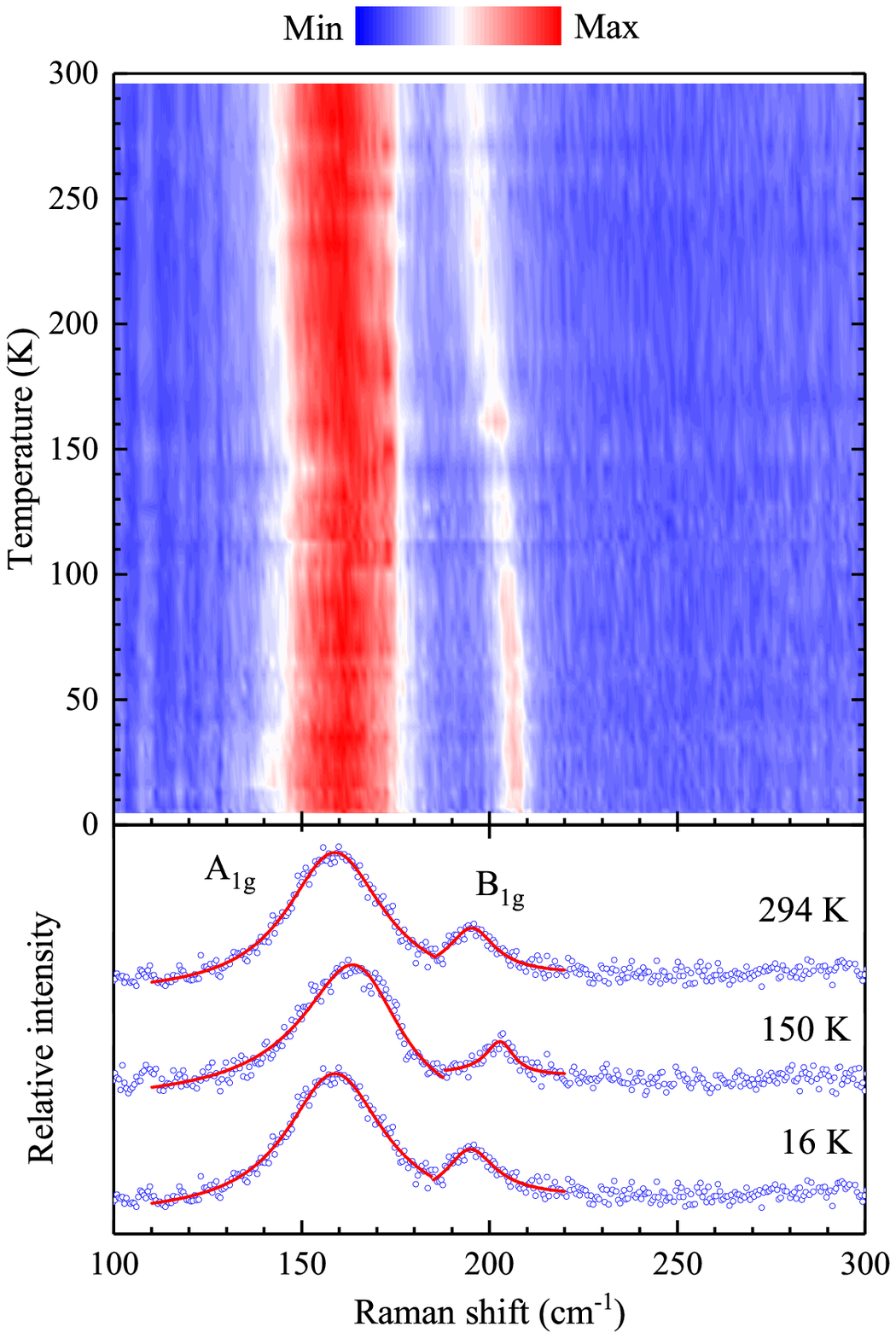}
	\setlength{\abovecaptionskip}{0.cm}
	\caption{(a) Mapping of the temperature-dependent Raman spectra of FeTe$_{0.5}$Se$_{0.5}$ from room temperature down to 4 K. The spectrum intensity denoted by different colors is presented at the top of the figure. The temperature evolution of the two phonon mode, the ${A}_{1g}$ and ${B}_{1g}$ mode, can be observed clearly. (b) Representative temperature-dependent Raman spectra of FeTe$_{0.5}$Se$_{0.5}$ at different temperatures. For clarity, three representative curves are displayed in this figure. The open circles and solid lines denote experimental data points and Fano fitting curves, respectively.}
\end{figure}

\subsection{Phonon spectra in the superconducting and normal state}

To further investigate the superconducting nature of FeTe$_{0.5}$Se$_{0.5}$, we systematically carried out the temperature-dependent Raman scattering measurements. The results are shown in Fig. 6. Two phonon modes, ${A}_{1g}$ and ${B}_{1g}$, were observed in the temperature-dependent mapping of Fig. 6(a). The frequencies of the ${A}_{1g}$ and ${B}_{1g}$ phonon mode at room temperature are about 158 and 192 cm$^{-1}$, respectively. They are close to the earlier reported values\cite{18,38}. The linewidth of the ${A}_{1g}$ mode is significantly larger than that of the ${B}_{1g}$ mode. Both phonon modes show a hardening behavior with decreasing temperature. The hardening modes were commonly reported in FeTe$_{x}$Se$_{1-x}$ and were thought to be caused by the temperature-induced lattice contraction\cite{18}. Surprisingly, compared with the generally hardening trend, there apparently is a slight softening behavior of the ${A}_{1g}$ mode below 40 K. This softening phonon mode will be analyzed and discussed in detail later. The representative Raman spectra at different temperatures are shown in Fig. 6(b). A marginally asymmetric phonon line shape emerges in the high-temperature range and becomes more notable at around 150 K. When the temperature decreases below 100 K, the symmetry of phonon peaks is profoundly refined. Taking this asymmetric effect into account, we used Fano function\cite{39,40} to fit the phonon spectra:
\begin{equation}
I(\omega)={I}_{0}\frac{(\frac{x-{\omega}_{0}}{\gamma}+q)^2}{1+(\frac{x-{\omega}_{0}}{\gamma})^2}+{I}_{b}(\omega)
\end{equation}
Where ${\omega}_{0}$ is the bare frequency, $q$ is the parameter quantitatively characterizing asymmetry, $\gamma$ is the phonon linewidth, ${I}_{0}$ is the parameter of intensity, and ${I}_{b}(\omega)$ denotes the background. The asymmetric lineshape turns back into Lorentzian lineshape when ${q}{\rightarrow}{0}$. The Fano fitting matches well with our experimental data. 

\begin{figure*}[t]
	\includegraphics[width=2\columnwidth]{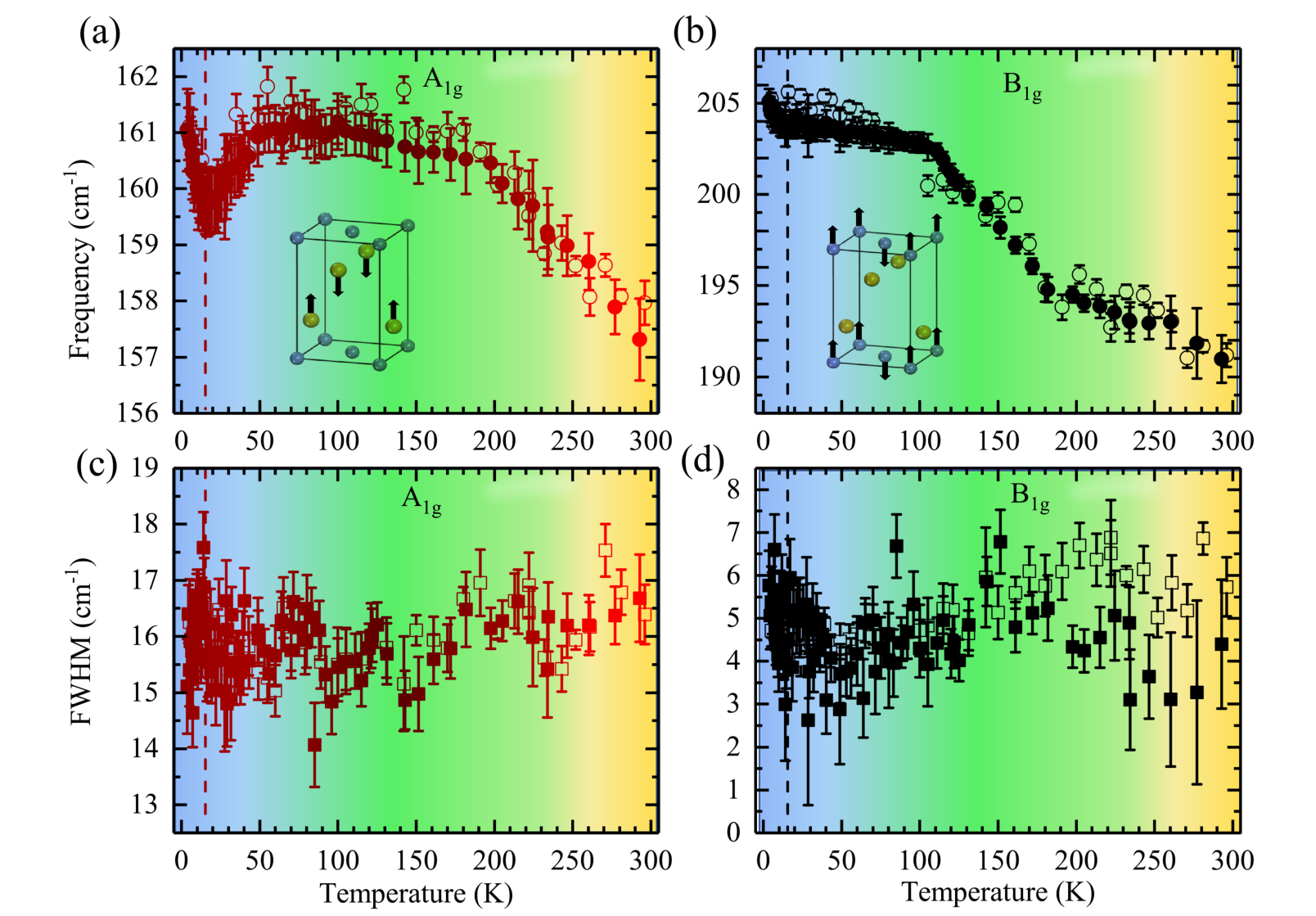}
	\setlength{\abovecaptionskip}{0.cm}
	\caption{(a)-(d) Temperature dependence of the phonon frequency and linewidth in FeTe$_{0.5}$Se$_{0.5}$. The crystal's point group symmetry probes the ${A}_{1g}$ and ${B}_{1g}$ channel. The inset graphs in (a) and (b) show the schematic illustrations for the ${A}_{1g}$ and ${B}_{1g}$ polarization mode. The blue atoms denote the Fe atoms, while the yellow atoms are the Te/Se atoms. The phonon frequency and linewidth of the ${A}_{1g}$ and ${B}_{1g}$ mode are displayed by the red and black symbols, respectively. The first and second rounds of the measurements are presented by the open and solid symbols, respectively. The four graphs are roughly divided into three parts using three different colors.}
\end{figure*}

The detailed temperature evolution of the frequency and linewidth for the ${A}_{1g}$ and ${B}_{1g}$ mode using Fano fitting is shown in Fig. 7. Considering that the ${A}_{1g}$ mode has a relatively small frequency change compared with its large phonon linewidth, we carefully designed and performed two-round Raman scattering experiments to ensure the consistency and accuracy of our results. We tried to measure the Raman spectra as detailed as possible in the low-temperature range which we most care about (4-50 K), with the temperature step less than 0.5 K. In the temperature range of 50-120 K, the temperature step was set to be about 1-3 K, while at higher temperatures this value was set to be around 5-10 K. The results of the first and second measurements match well with each other. Taking a brief look at the graph, we find a blueshift of the ${A}_{1g}$ and ${B}_{1g}$ mode when the temperature is decreased, as expected from the lattice contraction\cite{18}. The ${B}_{1g}$ mode shows a larger hardening behavior than the ${A}_{1g}$ mode, which can be attributed to the influence of the local spin state of Fe atoms\cite{17}. Upon cooling, the linewidth of the ${B}_{1g}$ mode generally narrows, with some abnormal changes showing up at around 250 and 40 K. The narrowing effect is well acknowledged because the linewidth is inversely proportional to the phonon lifetime\cite{41,42}. As the temperature cools down, the phonon lifetime increases, and the linewidth decreases. Yet, the narrowing of ${A}_{1g}$ mode is almost undecipherable.

Nevertheless, the ${A}_{1g}$ mode softens evidently at around 40 K, followed by the sign reversal of ${R}_{H}$ mentioned above. We also used the Lorentzian function to fit the phonon lineshape, which turns out to have almost the same softening behavior below 40 K. Below ${T}_{c}$, the ${A}_{1g}$ frequency rises again due to the self-energy effects\cite{68}.
The sudden reduction of the ${A}_{1g}$ phonon energy at 40 K, however, hasn't been reported previously in FeTe$_{0.5}$Se$_{0.5}$ (or near $x$ = 0.5) single crystals\cite{18}, probably because of the impurities or vacancies induced by imperfect growth conditions\cite{15,43}. Some similar softening anomalies have been found in FeTe\cite{17} and the interface of the single-layer FeSe film on substrates\cite{10}. Notably, the great enhancement of the ${T}_{c}$ in thin FeSe film can be interpreted as the consequence of the enhanced EPC interaction\cite{66,10}.  Moreover, previous calculations and ultra-fast dynamic experiments revealed that the ${A}_{1g}$ mode is the most substantial phonon mode in the EPC spectra of iron chalcogenides\cite{44,45}. Correspondingly, it is not surprising to find the softening behavior of the ${A}_{1g}$ mode in FeTe$_{0.5}$Se$_{0.5}$. Our results demonstrate the importance of the EPC on the superconductivity in iron chalcogenides.

\begin{figure}[t]
	\includegraphics[width=1\columnwidth]{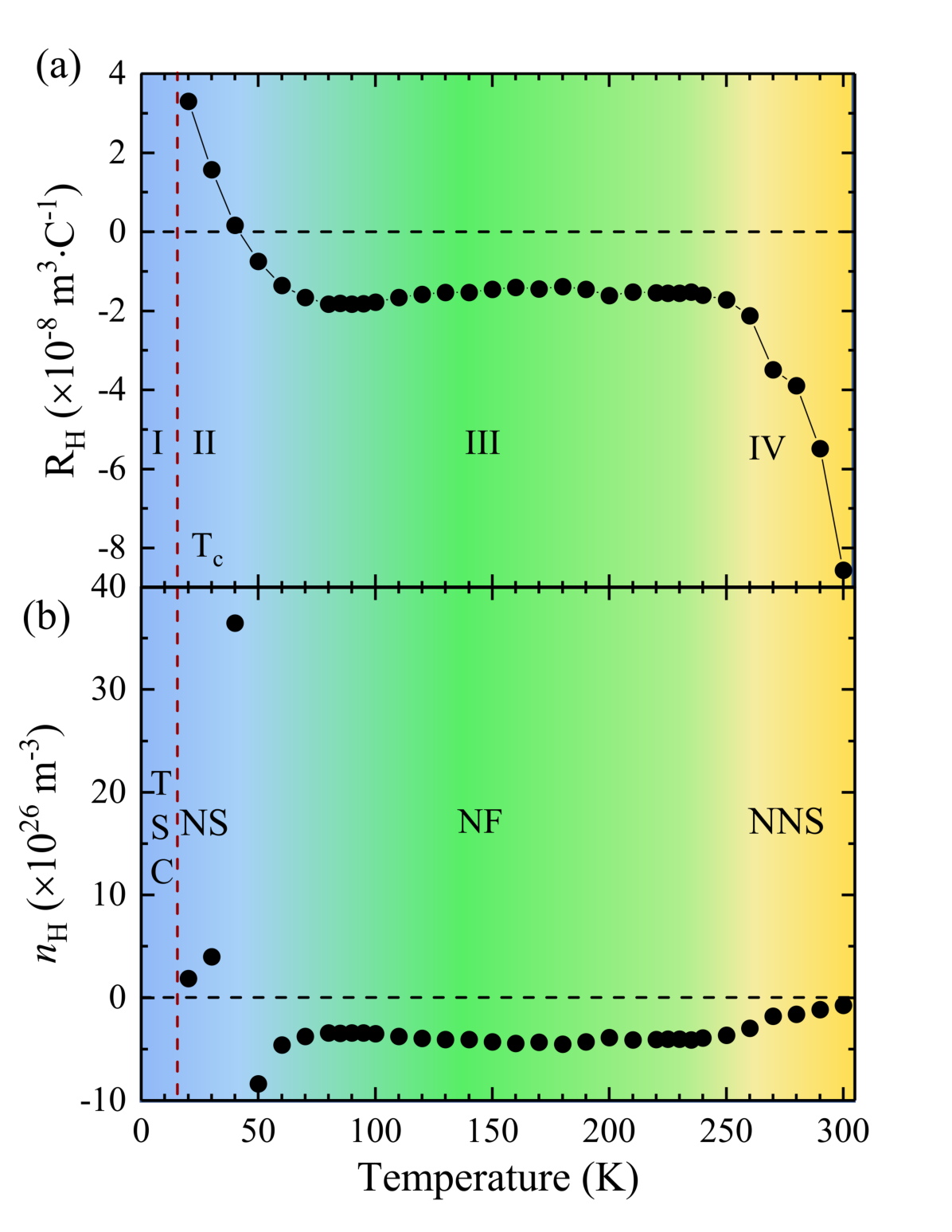}
	\setlength{\abovecaptionskip}{0.cm}
	\caption{Phase diagram of FeTe$_{0.5}$Se$_{0.5}$. Temperature dependence of the Hall coefficient (${R}_{H}$) (a) and the effective carrier concentration (${n}_{H}$) (b). The red dash line marks the ${T}_{c}$. The phase diagram can be divided into four parts using gradient ramp: (I) topological superconductivity (TSC), (II) nematic state (NS), (III) nematic fluctuations (NF), (IV) nonmetallic-normal state (NNS).}
\end{figure}

\subsection{Phase diagram for the superconducting, nematic, and normal state}

To better illustrate the electronic and structural dynamics, we summarize the temperature-dependent Hall coefficient and carrier concentration obtained from Eqs. (1) and (2) in Fig. 8. The graph can be divided into three parts, denoted as three different colors like Fig. 7, to show a schematic phase diagram. The effective carrier concentration at 20 K is obtained by Equation (5) [$P_{eff}$(20  K) = 1.65 $\times$ 10${{}^{23}}$ m$^{-3}$]. From 300 to 250 K (the yellow region IV), the sample shows a nonmetallic-normal state (NNS) for $d\rho/dT \leq 0$. The ${{R}_{H}}$ changes drastically in this temperature range. Below 250 K (the green region III),  a probable nematic fluctuations (NF) state is produced\cite{45}. At this time, the ${{R}_{H}}$ and ${n}_{H}$ suddenly turn to be almost constant. Simultaneously, an abnormality of the ${B}_{1g}$ linewidth seems to appear. Below 40 K (the blue region III), the nematic state (NS) emerges due to the Fermi surface reconstruction. With further cooling (the blue region on the left side of the red dash line), topological superconductivity (TSC) is induced by the band inversion of Fe$-{3d}$ and Te/Se$-{p}$ orbits near the hole pockets\cite{11,12}. The critical magnetic field measure for supporting the topological superconductivity in studied FeTe$_{0.5}$Se$_{0.5}$ was recently given by Yuan and Chen\cite{yuan}.

As seen from the schematic phase diagram, the topological superconductivity emerges right after the phonon softening and resides in the region of nematic order. It is natural to ask: what is the joint driving force behind these phenomena. The earlier widely accepted theory is that the spin fluctuations give rise to both nematicity and superconductivity\cite{5,46}.  
However, the effect of phonons, especially the ${A}_{1g}$ mode, cannot be simply omitted\cite{9,10,66}. In iron chalcogenides, the ${A}_{1g}$ mode would induce huge fluctuations through spin-phonon coupling\cite{47,67}. The fluctuations-enhanced EPC can be accountable for the experimentally observed ${T}_{c}$ of iron chalcogenides\cite{9,47}.
That explains why the ${A}_{1g}$ mode softens concomitantly with the nematic transition and becomes the softest at ${T}_{c}$. Our results provide a prospective approach to understanding the interrelationship between the spin fluctuations and EPC through transport and vibrational properties.

\section{CONCLUSIONS}
In summary, we have found the nematic transition of FeTe$_{0.5}$Se$_{0.5}$ followed by the topological superconductivity with strong spin-phonon interaction from both the electrical transport and Raman scattering experiments. At around 250 K, a probable fluctuation state shows up with the explicit changes of the Hall coefficient, carrier concentration, and phonon linewidth. As the temperature is decreased to about 40 K, the compound undergoes a clear nematic transition, which is confirmed by the sign reversal of the Hall coefficient and the concurrent softening of the ${A}_{1g}$ phonon mode. However, unlike other Se-doping rate samples such as FeTe$_{x}$Se$_{1-x}$, FeTe$_{0.5}$Se$_{0.5}$ does not exhibit an evident resistivity upturn at the nematic transition. That is probably due to the relatively small amplitude of nematicity compared with the extensive background. The softening ${A}_{1g}$ mode at nematicity indicates a muscular spin-phonon interaction and an enhanced electron-phonon coupling strength in this superconductor. As the temperature is decreased down to 20 K, the nonlinear Hall conductivity and linear magnetoresistance are observed. Both corroborate the topological superconductivity on the surface. The high-mobility nontrivial topological surface state has been identified successfully from the bulk state through the anomalous transport properties. Our results shed light on understanding the multiple electronic phases and the spin-phonon interaction in iron-based superconductors.

\section{ACKNOWLEDGEMENTS}
This work was funded through the National Key R\&D Program of China (Grant No. 2018YFA0305900) at HPSTAR, the Shenzhen Science and Technology Program (Grant No. KQTD20200820113045081), and the Basic Research Program of Shenzhen (Grant No. JCYJ20200109112810241). The work at BNL was supported by the US Department of Energy's the Office of Basic Energy Sciences with contract No. DOE-SC0012704.

\end{document}